\documentclass[12pt]{JHEP3}%\documentclass[12pt]{article}%JHEP3} %\input epsf.tex

\usepackage{epsfig}
\usepackage{amsmath,amssymb}

\newcommand{\CN}{{\mathbb{N}}}
\newcommand{\nn}{{\mathfrak{n}}}

\newcommand{\Tr}{{\rm Tr}}
\newcommand{\tr}{{\rm tr}}

\newcommand{\be}{\begin{equation} }
\newcommand{\ee}{\end{equation} }
\newcommand{\ba}{\begin{eqnarray}}
\newcommand{\ea}{\end{eqnarray}}

\preprint{KIAS-P04022
\\ UTIG-05-04\\ hep-th/0405244}

\title{Chern-Simons Solitons,
Chiral Model, and (affine) Toda Model \\
on Noncommutative Space}

\author{Ki-Myeong Lee \\ Korean Institute for Advanced Study, Seoul 130-722, Korea \\
${}^\dagger $Physics Department,University of Texas at Austin,
Texas 78712, USA
\\
Email: klee@kias.re.kr \\
${}^\dagger $({\rm until the end of May, 2004})}

\abstract{We consider the  Dunne-Jackiw-Pi-Trugenberger model of a
$U(N)$ Chern-Simons gauge theory coupled to  a nonrelativistic
complex adjoint matter on noncommutative space. Soliton
configurations of this model  are related the solutions of the
chiral model on noncommutative plane. A generalized Uhlenbeck's
uniton method for the chiral model on noncommutative space
provides explicit Chern-Simons solitons. Fundamental solitons in
the $U(1)$ gauge theory are shaped as rings of charge $\nn$ and
spin $-\nn$ where the Chern-Simons level $\nn$ should be an
integer upon quantization. Toda and Liouville  models are
generalized to noncommutative plane and the solutions are provided
by the uniton method. We also define  affine Toda and sine-Gordon
models on noncommutative plane. Finally the first order moduli
space dynamics of Chern-Simons solitons is shown to be trivial. }

\begin{document}
\baselineskip 18pt

\section{Introduction and Conclusion}

There has been considerable interest in finding solitons  on
noncommutative space. (For reviews, see
Refs.~\cite{Douglas:2001ba,Harvey:2001yn,Szabo:2001kg} for
example.) Classically well-known solitons, like instantons,
magnetic monopoles, vortices on commutative space have found also
on the theories defined on noncommutative plane. In addition,
there are certain solitons whose existence is possible only on
noncommutative plane. Some are intrinsically defined only on
noncommutative plane~\cite{Gopakumar:2000zd}. Others are extension
of solitons to the $U(1)$ gauge theories~\cite{Nekrasov:1998ss}.

Sometime ago Dunne, Jackiw, Pi and Trugenberger (DJPT) have
studied solitons of zero energy in a Chern-Simons gauge theory
coupled to nonrelativistic complex matter field in  adjoint
representation~\cite{Dunne:1990qe,Dunne:1992hq}.   Once a
consistent ansatz is made for the field configuration,  the
self-dual  equations for these Chern-Simons solitons can be
reduced to the equations for the Liouville and Toda models whose
solutions are known~\cite{Dunne:1990qe}.  In addition the
equations for the affine Toda and sinh-Gordon models were shown to
be derivable from a consistent ansatz.  In
Ref.~\cite{Dunne:1992hq} all Chern-Simons  soliton configurations
have been found to be related to the solutions of the chiral
theory, which can be found by Uhlenbeck's uniton
method\cite{uhlenbeck,wood}. In addition the solutions for the
Toda model were recovered  by the uniton method.

In this work we investigate the extension of the DJPT model to
noncommutative space, which is possible only for $U(N)$ gauge
group. We find that the correspondence between  the solutions of
the  self-dual equation and the chiral model works equally well on
noncommutative plane. The uniton method works as well with some
caveat. Interestingly the Chern-Simons solitons on $U(1)$ gauge
theory are made of fundamental `ring' shape configurations of
charge $\nn$ and spin $-\nn$ with  the Chern-Simons level $\nn$. A
consistent ansatz for (affine) Toda model can be found also on
noncommutative space, even though the first order equations cannot
be reduced to second order equations as  on commutative space. Our
equations can be regarded as a generalization of Liouville,
sinh-Gordon, and (affine) Toda models to noncommutative space.
Using the uniton method, we find also the solution of the
generalized Liouville and Toda equations on noncommutative plane.
Note that the explicit solutions of Toda models on commutative
space cannot be translated easily to those on noncommutative
space. In addition, we show that the moduli space dynamics of
Chern-Simons solitons is trivial.

The Chern-Simons solitons of
Dunne-Jackiw-Pi-Trugenberger~\cite{Dunne:1990qe,Dunne:1992hq}
belong to a large class of self-dual solitons appearing in
Chern-Simons theories coupled to matter fields, where the global
or local U(1) symmetry is essential for their existence. (See for
reviews \cite{Dunne:ai}.) They could come in varieties as vortices
in asymmetric phase, q-balls or q-balls with vortices  in
symmetric phases. The DJPT solitons can be regarded as the
nonrelativistic  limit of q-balls with vortices in middle of
certain class of the relativistic theories. What is interesting
about DJPT model is that the lowest or zero energy solutions
satisfy the self-dual equations which is dimensional reduction of
the self-dual Yang-Mills equation on space with (2,2) signature.
Indeed they made a correspondence between the self-dual equations
with the equations for the chiral model~\cite{Dunne:1990qe}, and
in Ref.~\cite{Dunne:1992hq} the complete solutions are found by
Uhlenbeck's uniton method~\cite{uhlenbeck,wood}.

In addition, they found that upon choosing a consistent ansatz,
the self-dual equations are reduced to Toda or affine Toda
equations. For that of $SU(2)$, these equations can be shown to
contain Liouville and sinh-Gordon equations, respectively. In DJPT
papers, they  have provided a version of the general solution of
$SU(N)$ Toda equation which can be also obtained from the uniton
method.  Their work can be also understood in the context of
Ward's conjecture~\cite{ward2}, which associates all integrable
models to the   dimensional reduction of self-dual Yang-Mills
equations on flat four dimensional space with $(4,0)$ or $(2,2)$
signature.

The solitonic physics on noncommutative space has been considered
recently.  New features of solitons can appear on noncommutative
space. Some solitons which can collapse to points and disappear on
ordinary space reaches a finite size on noncommutative space. On
noncommutative space their moduli space becomes complete in short
distance. Instantons in the $U(N)$ gauge theory on noncommutative
four space provides a prototype of noncommutative
example~\cite{Nekrasov:1998ss}.

A pure neutral scalar field theory could have nontrivial solitons
on noncommutative space~\cite{Gopakumar:2000zd}. In large
noncommutative limit, a projection operator on Hilbert space
defined by the noncommutative spatial coordinates  becomes a
solitonic configuration.  The dynamics of these solitons are
explored in Ref.~\cite{Gopakumar:2001yw,Lindstrom:2000kh}. These
operators will play a role in the physics of Chern-Simons
solitons. The chiral models on noncommutative space has been
explored by the dressing methods~\cite{Lechtenfeld:2001uq}, where
some of solutions are explored in detail. Our approach here by
unitons is somewhat different from their approach and probably
provides all soliton solutions in one setting.  Our work on
unitons on noncommutative space  are more closely related to the
self-dual solutions on $CP_N$  and grassmannian models, which has
been explored in many
directions~\cite{Lee:2000ey,Furuta:2002ty,Otsu:2003fq}.

Our generalization of the DJPT model to noncommutative space is
straightforward. The Chern-Simons solitons appear as a lump on
noncommutative plane with integer global U(N) charge and integer
spin, whose minimum value is $\nn$ and $-\nn$ respectively with
the Chern-Simons level $\nn$, which should be integer upon
quantization~\cite{Deser:vy,Bak:2001ze,Nair:2001rt}. The general
Chern-Simons solitons are composite of finite number of
fundamental solitons of minimum charge. When the gauge group is
$U(1)$, each fundamental soliton has two moduli parameters
accounting for its position. When the gauge group is larger, there
are also scale and internal orientation parameter, exactly like
instantons on noncommutative four dimensional space.

%There is a $U(N)$ universality on noncommutative space, which
%means that the $U(N)$ gauge theory can be rewritten as the $U(1)$
%gauge theory~\cite{Gross:2000ss,Bak:2000zm}. If the physics
%explores large distances, one could say the boundary condition at
%spatial infinity fixes the gauge group. In our DJPT model, all
%f%ield configurations fall off to zero at spatial boundary quickly,
%implying that the $U(N)$ universality works here. The implication
%of this on the Chern-Simons solitons is quite interesting and
%worth further investigation.

The configurations for Chern-Simon solitons on noncommutative
space cannot be found easily by trying to translate the explicit
solutions of Toda equations on commutative space. Here we relate
the self-dual equations for Chern-Simon solitons on noncommutative
space to the equation of $U(N)$ chiral model on noncommutative
space. Fortunately the chiral equation can be solved by the uniton
method. On commutative space the unitons provide the most general
solution. While it is not yet proven, this seems to be case also
on noncommutative space. Thus the uniton method seem to provide
general Chern-Simons soliton configurations.

In addition, the reduction to (affine) Toda model by a consistent
ansatz works equally  well on noncommutative space once the gauge
group is $U(N)$ and the ordering of operators are carefully taken
care of. This leads to a generalization of (affine) Toda model to
noncommutative space. For $U(2)$ case, the self-dual equation with
a proper consistent ansatz leads to Liouville and sinh-Gordon
equations. In our generalization, we still have coupled first
order equations on noncommutative space, which cannot be reduced
to coupled second order equations in general. In recent works on
integrable models on noncommutative space~\cite{integrable},  the
coupled second order equations are generalized to noncommutative
plane. It remains to be seen how two generalizations to
noncommutative space are related each other. While our work is
focused on Euclidean noncommutative plane, it is straightforward
to reduce the self-dual Yang-Mills equations with (2,2) signature
along the Minkowski noncommutative plane. Again the equations
obtained from this angle would be coupled first order equations.

Finally we study the moduli space dynamics of Chern-Simons
solitons briefly, which is first order in time-derivative. There
are moduli parameters for Chern-Simons solitons. The first order
Lagrangian would lead  to a first order Lagrangian on moduli
space.  It turns out to be trivial as the first order induced
kinetic term on moduli space vanishes. This is consistent with the
fact that the total angular momentum of many Chern-Simons solitons
is independent of their position and so no nontrivial interaction
required for  spin-statistics theorem is expected. There are
several works on moduli space dynamics of solitons in $CP_N$ and
chiral models on noncommutative plane, which as some relevance to
our work
here~\cite{Gopakumar:2001yw,Lindstrom:2000kh,Lechtenfeld:2001uq,Furuta:2002ty,Wolf:2002jw}.
They are basically looking at the second order kinetic term, which
leads to the smooth moduli space metric of identical particles.

In addition, it would be interesting to find out  nontrivial
time-dependent solutions of Chern-Simons solitons which is not
captured in the moduli space dynamics. The $U(1)$ Chern-Simons
theory coupled to a nonrelativistic matter field in fundamental
representation has been studied~\cite{Bak:2001sg} and some
time-dependent solutions has been found~\cite{Hadasz:2003yx}. A
pursuit along this direction may be profitable.

The plan of this work is as follows. In Sec.2, we introduce the
DJPT model on two dimensional noncommutative space, and explore
some basic properties. In Sec.3, we relate the self-dual equations
for Chern-Simons solitons to the equation of the chiral model. We
review and extend the uniton method to  noncommutative space.  We
also study some basic properties of Chern-Simons soliton
solutions. In Sec.4, we introduce a consistent ansatz and obtain
(affine) Toda model on noncommutative space. We also study a
explicit solution obtained by the uniton method. Finally in Sec.5,
we show that the first order moduli space dynamics of the
Chern-Simons solitons is trivial.

\section{DJPT Theory on Noncommutative
Space}

Our space is a two dimensional noncommutative space with
noncommuting  coordinates $(x,y)= (x^1,x^2)$ satisfying
\be [x,y] = i\theta \; ,\ee
where $\theta>0$ without lose of generality. The complex
coordinate variables are $z= (x+iy)/2$ and $\bar{z}=(x-y)/2$. The
creation and annihilation operators are defined as
\be a = \frac{x+iy}{\sqrt{2\theta}} = \sqrt{\frac{2}{\theta}} \;
z,\; \bar{a} = \frac{x-iy}{\sqrt{2\theta}} =
\sqrt{\frac{2}{\theta}} \; \bar{z} \; , \ee
which satisfy  $ [a,\bar{a}]=1$. The Hilbert space of harmonic
oscillator is given by $\{ |n\rangle,n=0,1,2,...\}$ such that
$a|0\rangle=0$ and $|n\rangle = \bar{a}^n/\sqrt{n!} |0\rangle$.
The three dimensional space-time coordinates are
$(x^0,x^1,x^2)=(t,x,y)$. Any field  $\phi(x,y)$ on noncommutative
space   becomes an operator on the Hilbert space. The space
integration becomes the trace over this Hilbert space $\int d^2 x
= 2\pi \theta\; \Tr $.  The spatial derivatives $\partial_i \phi $
can be represented by operators $
\partial_x \phi= i[y,\phi]/\theta$, $\partial_y \phi = -i[x,\phi]
/\theta $. In the complex variables
\be \partial_+ = \partial_x+i\partial_y = \partial_{\bar{z}}
=\frac{2}{\theta} [z,\;] ,  \;\; \partial_- =
\partial_x-i\partial_y = \partial_{z} =\frac{2}{\theta}
[\bar{z},\;] \; . \ee

The antihermitian gauge fields $A_\mu = A_t, A_x, A_y$ for the
$U(N)$ gauge group define a covariant derivative on the scalar
field $\phi$ in the adjoint representation via $D_i\phi =
\partial_i \phi +[A_i,\phi]$, where $\phi$ and $A_i$ are $N$ by $N$ matrices.
On our noncommutative plane, the covariant derivatives can be
rewritten as
\be D_x \phi =  \frac{i}{\theta} [Y,\phi], \;\;
D_y\phi=\frac{-i}{\theta}[X,\phi]  \; , \ee
where the covariant position operators are
\be X = x + i\theta A_y,\;\; Y= y-i\theta A_x\; . \ee
The covariant complex position operator and its conjugate are $
Z=(X+iY)/2$ and $\bar{Z}=(X-iY)/2$.  The complex covariant
derivatives are
\be D_+= \partial_+ + A_+ =\frac{2}{\theta}[Z,\;\;], \;\;
D_-=\partial_-+ A_- = \frac{2}{\theta}[\bar{Z},\;\;]\; , \ee
where  $A_\pm = A_x\pm i A_y$. In complex coordinates, the  field
strength becomes
\be  F_{+-} = \partial_+ A_--\partial_-A_++[A_+,A_-]= -2i F_{xy}\;
. \ee

The theory we consider here is the $U(N)$ Chern-Simons gauge
theory coupled to a nonrelativistic complex bosonic  field $\phi$
in the adjoint representation, which is a simple generalization of
the $SU(N)$ DJPT model. (A pair of hermitan fields in the adjoint
representation are combined to a complex field in the adjoint
presentation.) The gauge field $A_\mu$ is $N$ by $N$ antihermitian
matrices and the matter field $\phi$ is $N$ by $N$ complex matrix.
Its Lagrangian is
\be L  = -\frac{\nn}{4\pi } \int d^2 x \left\{
\epsilon^{\mu\nu\rho}\; \tr \left( A_\mu
\partial_\nu A_\rho + \frac{2}{3}A_\mu A_\nu A_\rho \right)  -i \tr \left(
 \phi^\dagger D_0 \phi\right) \right\}  - H\; , \label{Lagrang}\ee
where $D_\mu \phi= \partial_\mu \phi + [A_\mu,\phi]$ and $\tr$ is
a trace over  $N\times N$ matrices.

Up on the quantization of the Chern-Simons theory  on
noncommutative plane, the Chern-Simons level $\nn$ should be
quantized in integer for
consistency~\cite{Deser:vy,Bak:2001ze,Nair:2001rt} for all
naturall number  $N$.   Our choice of the scalar field
normalization is for convenience. In order to obtain the standard
kinetic energy for the matter field, we need to change the
normalization of the $\phi$ field, with possible charge
conjugation. The conserved energy is
\be H =  \frac{\pi |\nn|}{m} \int d^2 x  \;  \tr \biggl( 2
D_i\phi^\dagger D_i\phi - [\phi,\phi^\dagger]^2 \biggr)\; .
\label{energy} \ee
The mass parameter $m$ is positive. The potential is negative and
so attractive. With the standard normalization of the $\phi$ field
$m$ is the kinetic mass and the attractive interaction is
inversely proportional to $\nn$.

On noncommutative plane the allowed gauge group is $U(N)$ instead
of $SU(N)$. Under the $U(N)$ gauge transformation,
\be \phi\rightarrow U^\dagger(x) \phi U(x),\;  A_\mu \rightarrow
U^\dagger(x) A_\mu U(x) + U(x)\partial_\mu U(x)\, . \ee
The Gauss's law constraint is
\be F_{+-} = -2i F_{xy} = [\phi,\phi^\dagger]  \, .\label{gauss1}
\ee
There is a global $U(1)$ symmetry $\phi \rightarrow
e^{i\alpha}\phi$, whose conserved charge is
\be Q = \frac{\nn}{4\pi} \int d^2x \; \tr \phi\phi^\dagger \, ,
\label{charge1} \ee
which is expected to be quantized in integer.

To consider the lowest energy configuration for a given charge, we
 first note that
\be \tr(D_i\phi^\dagger D_i \phi) = \tr\biggl((D_+\phi^\dagger) (
D_- \phi)- i[\phi,\phi^\dagger]F_{12} \biggr) +i\epsilon_{ij}
\partial_i \tr ( \phi^\dagger D_j \phi) \, . \ee
With the Gauss law (\ref{gauss1}),  the conserved energy
(\ref{energy}) for any  localized configuration  becomes
\be H =  \frac{2\pi |\nn|}{m}  \int d^2x  \;  \tr
(D_+\phi^\dagger) (D_-\phi) \, , \ee
which is  nonnegative. Its minimum, zero energy, can be achieved
by  localized configurations satisfying the Gauss law constraint
(\ref{gauss1}) and the self-dual equation
\be  D_- \phi = 0\, . \label{selfdual}\ee
The physics of these zero energy configurations, Chern-Simons
solitons, on noncommutative plane is the main  concern here. Note
that the Gauss law (\ref{gauss1}) and  selfdual equation
(\ref{selfdual}) can be regarded as a dimensional reduction of the
selfdual Yang-Mills equation on four dimensional space of
signature (2,2).

Under the  infinitesimal rotation $\delta x^i = \epsilon_{ij} x^j
$ with the corresponding field transformation
\ba && \delta \phi = \frac{1}{2} \epsilon_{ij} (X^i(D_j \phi)
+(D_j \phi)X^i) \, ,\\
&& \delta A_i =   -\frac{1}{2}(X^i F_{12} +F_{12}X^i) \, ,\ea
leaves the action invariant modulo a gauge transformation. Note
that the symmetrized transformation of the field is essential  on
noncommutative plane. The corresponding conserved angular
momentum~\cite{Bak:2001ze} is
\be   J = \frac{\nn i}{16\pi}  \int d^2x \; \epsilon_{ij} \tr \;
X^i \biggl( (D_j \phi \phi^\dagger -\phi D_j \phi^\dagger) +
(\phi^\dagger D_j\phi-D_j\phi^\dagger \phi) \biggr)\, , \ee
which has a gauge-covariant density. The rotational symmetry group
is $U(1)$ on the noncommutative plane. The conserved linear
momentum can be obtained by noting that under shifting $X^i
\rightarrow X^i+a^i$, $J \rightarrow J + \epsilon_{ij} a^i P^j$.
For the self-dual configurations satisfying (\ref{gauss1}) and
(\ref{selfdual}), the above angular momentum becomes
\be J = \frac{\nn}{16\pi} \int d^2x \; \biggl\{ Z (\phi D_-
\phi^\dagger + D_-\phi^\dagger \phi) + \bar{Z}(D_+\phi
\phi^\dagger+ \phi^\dagger D_+\phi) \biggr\} \, .\ee
Integration by part leads to
\be J= -Q \, .\ee
The charge and angular momentum quantizations are consistent to
one another.

\section{Chiral Model and Unitons}

Similar to the commutative case, one can easily map  the Gauss law
and the self-dual equation to the field equation for the chiral
model of $U(N)$ group on noncommutative plane. As in
Ref.~\cite{Dunne:1990qe}, we  introduce a new `auxiliary gauge
field'
\be {\cal A}_+= A_+ - \phi,\;\; {\cal A}_-= A_- + \phi^\dagger\, ,
\ee
which has zero field strength
\be {\cal F}_{+-} = \partial_+ {\cal A}_- - \partial_- {\cal A}_+
+ [{\cal A}_+, {\cal A}_-] = 0 \ee
due to the Gauss law (\ref{gauss1}) and the self-dual equation
(\ref{selfdual}). (One can introduce arbitrary phase difference
between $A_+$ and $\phi$ of ${\cal A}_+$ field, which corresponds
to the spatial $U(1)$ and global $U(1)$ rotations, and so
immatiarial.) Thus the gauge field ${\cal A}_\pm$ should be pure
gauge which we choose to be zero. (Later we will choose a gauge
transformation of this $A_+=\phi$ configuration, leading to
nonzero ${\cal A}_+$.) Thus,
\be  A_+=\phi\, , \;\; A_- =-\phi^\dagger \, , \label{aphieq} \ee
which solves the Gauss law. Now the self-dual equation
(\ref{selfdual}) and its hermitan conjugate become
\be \partial_ -\phi =[\phi^\dagger,\phi],\;\; \partial_+
\phi^\dagger  = [\phi^\dagger,\phi]\, , \ee
which are  equivalent to two equations
\ba && \partial_+ \phi^\dagger -\partial_-\phi  = 0 , \label{first}\\
&& \partial_+ \phi^\dagger +\partial_-\phi +2[\phi, \phi^\dagger]=
0 \, . \label{second} \ea
Introducing another auxiliary gauge field $C_+=2\phi$ and
$C_-=-2\phi^\dagger $, the second equation (\ref{second}) becomes
the zero field strength equation for $C_\pm$. This simplies that
$C_\pm$ are pure gauge or
\be \phi = \frac{1}{2} h^{-1}\partial_+ h\, ,\;\; \phi^\dagger  =
- \frac{1}{2} h^{-1}
\partial_- h \, ,\label{phi-heq}\ee
for a map  $h(x)$ from the noncommutative plane to $U(N)$ group.
The first equation (\ref{first}) implies that $h$ satisfies
\be \partial_+(h^{-1}\partial_-h)+ \partial_-(h^{-1}\partial_+h) =
0 \, , \label{chiraleq} \ee
which is the equation for the $U(N)$ chiral model on
noncommutative plane.

This chiral equation can be derived from the stationary point of
the nonnegative energy functional
\be {\cal H}_{chiral} = -\frac{1}{2} \int d^2x \; \tr ( h^{-1}
\partial_i h)^2 \, .
\label{hchiral}\ee
As the chiral field $h$ is related to the complex scalar field
$\phi$ by the relation (\ref{phi-heq}), the conserved $U(1)$
charge (\ref{charge1}) of the Chern-Simons theory becomes
\be Q = -J= \frac{\nn}{8\pi} {\cal H}_{chiral} \label{hqjeq} \ee

On the commutative plane there exists an algorithm to find all
solutions of the chiral model which have the finite energy ${\cal
H}_{chiral}$ by using the so-called unitons~\cite{uhlenbeck,wood}.
One key aspect of the theorem is a method to generate new
solutions from old ones, by the so-called `the addition of a
uniton' procedure. It  works equally well on noncommutative plane
as we will see now. For a given solution $h_0$ of the chiral model
equation, one defines a `uniton' factor with respect to $h_0$ to
be $2p-1$ with a hermitian projection operator $p$  which
satisfies two equations,
\be (1-p)(\partial_+p + A_{+}^0p)=0,\;\; pA^0_+(1-p) = 0 \ee
where $A_\pm^0= \frac{1}{2} h_0^{-1}\partial_\pm h_0 $. Then the
operator $h=h_0(1-p)$ is also a solution of the chiral model
equation because for such $h_0$ and $p$,
\be h^{-1}\partial_+ h =2(\partial_+p +A_+^0),
\;\; h^{-1}\partial_- h = 2(-\partial_-p + A_-^0) \ee

 Let us now put forward a conjecture on noncommutative plane
which is a simple generaliaztion of the theorem due to
K.Uhlenbeck~\cite{uhlenbeck} and refined by J.C. Wood~\cite{wood}.
We generalize the  version appeared in Ref.~\cite{Dunne:1992hq} to
the U(N) chiral model   on noncommutative plane, where the
operator ordering is crucial matter.

\noindent{\bf Conjecture I:}  Every  solution $h$ of the $U(N)$
chiral model equation (\ref{chiraleq}) with finite energy
(\ref{hchiral}) on noncommutative plane can be uniquely factorized
as a product
\be h = (2p_1-1)(2p_2-1)...(2p_m-1) \label{heqfac} \ee
with $m\le N-1$  where  (a)  each $p_i$ is a hermitian $N\times N$
projection operator on noncommutative plane,  and \; (b) with a
definition  $h_j=(2p_1-1)(2p_2-1)...(2p_j-1)$, $2p_{j+1}+1$ is a
uniton with respect to $h_j$ for all $j=1,...,m-1$.

We do not know whether above conjecture is true or not. It remains
to be explored further. There exists also a theorem by
G.~Valli~\cite{valli} on the value of the energy (\ref{hchiral})
for a solution $h$ of the chiral model on commutative plane. Its
generalization to noncommutative space becomes as follows:

\noindent{\bf Conjecture II:} Let $h$ be a solution of the chiral
model equation (\ref{chiraleq}) on noncommutative plane, which can
be expressed as in Eq.~(\ref{heqfac}). Then the conserved energy
(\ref{hchiral}) takes a nonnegative integer mulitiple of $8\pi$.

This theorem holds for all configurations we will study later and
is  expected to hold on noncommutative space. Thus the charge and
spin of Chern-Simons solitons would be positive integer multiples
of $\nn$ and $-\nn$. We could say  that a general Chern-Simons
solitons are  made of integer number of fundamental Chern-Simons
solitons of charge $\nn$ and spin $-\nn$.

To quadratic order, elementary particles describes by the field
$\phi$ carries unit charge and so one could regard the fundamental
Chern-Simons solitons as a composite of $\nn$ elementary particles
bound together by an attractive force of strength $1/\nn$.  While
we expect elementary particles are anyons with nontrivial
spin-statistics, even may with nonabelian braid group, fundamental
Chern-Simons solitons carry integer spin upon quantization as
$\nn$ is quantized in integer.

The weak coupling limit is the large $\nn$ limit, and one can see
that the quartic attractive interaction becomes weak as it is
proportional to $1/\nn$ after a proper normalization of the scalar
field. Thus our classical picture of Chern-Simons solitons becomes
better for large $\nn$. While it is in principle possible for our
Chern-Simons solitons to have purely nonabelian statistics, we
will see later that  no such interaction is apparent on first
order dynamics on the space of moduli parameter.

There are studies of some explicit  chiral solutions by using
unitons~\cite{zak,ward} on commutative plane, which would have
also application on noncommutative space.  More recently, the
solutions of the chiral model on noncommutative plane  have been
explored by the dressing method \cite{Lechtenfeld:2001uq}. Here
our goal is to exploring the role of unitons on both Chern-Simons
solitons and the solutions of chiral model on noncommutative
plane. The understanding obtained from the soliton solutions on
$CP_N$ model plays a key role here~\cite{Lee:2000ey,Otsu:2003fq}.

Assuming above two conjectures are true, let us explore their
imiplications on Chern-Simons solitons.  First let us consider the
simplest case,  the the chiral and DJPT model of $U(1)$ group on
noncommutative plane. Note that these models with $U(1)$ group do
not have nontrivial solutions on commutative plane. The uniton
solution would be
\be h = (2p-1)  \ee
such that
\be (1-p)\partial_+ p = 0 \label{peq}\ee
which is equivalent to $(1-p) ap = 0$.  The hermitan projection
operator satisfying $p^2=p$ has been extensively studied in the
study of solitons on noncommutative plane~\cite{Gopakumar:2000zd}.
The projection operator for $K$ number of solitons at  positions
$z_r=(x_r+iy_r)/2$ is given by
\be P_K = \sum_{r,s=1}^K |z_r\rangle \langle z_r|z_s\rangle^{-1}
\langle z_s| \label{pk} \ee
with
\be |z_r\rangle = \exp\left( -\frac{|z_r|^2}{\theta} +
\sqrt{\frac{2}{\theta}} \; z_r a^\dagger \right) |0 \rangle \ee
Note that $(z-z_r)|z_r\rangle=0$ for all $r$. Interestingly, the
above projection operator $P_K$ satisfies the equation
(\ref{peq}), implying $h=h^{-1}= 2P_K-1$ is the solution of the
chiral equation. For $p$ satisfying (\ref{peq}) and having finite
trace, the conserved Hamiltonian (\ref{hchiral}) becomes
\be  {\cal H}_{\rm chiral} = -8\pi\, \Tr [a,p][\bar{a},p] =
-8\pi\, \Tr (ap\bar{a}-pa\bar{a}) = 8\pi \, \Tr p \ee
Thus for $p=P_K$, the above quantity would be $8\pi K$, confirming
Conjecture II.

For such a solution of the chiral equation, the soliton solution
of the DJPT model would be given by $A_+=\phi=\partial_+ P_K$. The
total charge and angular momentum are $Q= -J= \nn K $. When
particles are all coming together the projection operator becomes
$P_K = \sum_{r=0}^{K-1}|r\rangle\langle r|$, and so the  magnetic
field of the corresponding Chern-Simons soliton becomes
 \be F_{+-} = [\phi,\phi^\dagger] = K\bigl(|K-1\rangle\langle K-1|
 -|K\rangle\langle K| \bigr) \ee
so that the net magnetic flux vanishes. Our Chern-Simons solitons
can be regarded as a nonrelativistic limit of  Q-balls with vortex
at center in the symmetric phase of the relativistic Chern-Simons
Higgs theory~\cite{Jackiw:1990pr,Lee:1990bd,Dunne:fs}. The above
field strength shows the remnant of this configuration. On the
right side the first term is the vortex contribution and the
second term is the Q-ball contribution. The angular momentum could
be regarded as the difference between the vortex contribution and
the Q-ball contribution. Our solutions describe  a composite of
$K$ number of fundamental Chern-Simons solitons of ring shape.

In the $U(N)$ theory, we consider a $N$ by $N'<N$ matrix
\be M= \left(\begin{array}{cccc} f_{11}(z) & f_{12}(z) & ...& f_{1N'}(z) \\
. & .& ...& . \\
f_{N1}(z)& f_{N2}(z)& ...& f_{NN'}(z)
\end{array}\right) \label{sunitonm} \ee
where $f_{ij}(z)$ are polynomials of $z$. With a projection
operator defined as
\be p = M \frac{1}{\bar{M}M} \bar{M} \label{sunitonp}\ee
It satisfies $(1-p)\partial_+ p = 0 =(\partial_+p)p$, implying
$h=2p-1$ satisfies the chiral equation. Thus the above projection
operator $p$ can be regarded as a single uniton. The $N$ by $N'$
matirx operator
\be \Phi = M (\bar{M}M)^{-1/2} \ee
satisfies $\Phi\Phi^\dagger = p $ and $\Phi^\dagger \Phi =
1_{N'N'}$. It can be regarded as an harmonic map from the
noncommutative plane to an Grassmann
manifold~\cite{Lee:2000ey,Furuta:2002ty,Otsu:2003fq}.

This contrasts to the commutative case where $f_{ij}$ can be
rational analytic functions of $z$. As we have seen that a single
uniton solution is closely related to the $CP_{N-1}$ or Grassmann
solitons on noncommutative plane, in which case the entries of $M$
being polynomial, not a rational function, is essential for
consistency~\cite{Lee:2000ey}. The point is that a rational
function has poles where it becomes not holomorphic due to
Dirac-delta function. It can be neglected on commutative space but
not on noncommutative space, and so the correspodning  projection
operator $p$ is not a uniton any more, and leads to un-quantized
topological charge.

The trace of the above $p$ is infinite. With $h=2p-1$, the chiral
energy becomes the topological charge for $\Phi$ as
\ba {\cal H}_{chiral} &=& 4\pi \theta\; \Tr \tr_N\partial_+ p
(1-p)\partial_-p \nonumber \\
&=& 4\pi \theta \Tr \tr_{N'} \partial_+ \bar{\Phi}
(1-p)\partial_-\Phi \nonumber \\
&=& 4\pi \theta \Tr \tr_{N'} \partial_+\left( \frac{1}{\bar{M}M}
\bar{M}\partial_-M\right) \ea
The argument in Ref.~\cite{Lee:2000ey} shows that the above
quantity becomes $8\pi$ times the highest  degree of the
polynomials appear in $M$.

Especially with $U(2)$ group, we consider a single uniton solution
defined by a 2 dimensional  vector
\be M =\left( \begin{array}{c} f(z) \\ c \end{array}\right) \ee
where $f(z)$ is the $K$-th order polynormial
\be f(z) = \prod_{r=1}^K (z-z_r) \ee
The corresponding projection operator becomes
\be p = \left(\begin{array}{c} f(z)\\ c \end{array}\right)
(\bar{f}f +|c|^2)^{-1} (\bar{f}(z) , \bar{c}) \ee
In the limit  $c\rightarrow 0$, this   projection operator
approaches
\be p\rightarrow \left( \begin{array}{cc} 1  & 0 \\ 0 &  P_K
\end{array}\right) \ee
with $P_K$ in Eq.~(\ref{pk}). The reason is that $P_K$ is only
projection operator such that $(z-z_r)P_K=0$ for all $r$. The
above arguement shows that  the chiral energy can be shown to be
$8\pi K$. For $K=1$, $f(z) = z-z_1$ and so  the chiral energy is
$8\pi $, and the charge and spin of the Chern-Simons soliton is
$Q=-J=\nn$, indicating that it is a fundamental one.

The counting of the parameters of fundamental soliton is a bit
subtle. On chiral model and $CP_N$ model, there are vacuum
parameters, which need to be specified. Those parameters has
infinite inertia and so does not change with time. The solutions
of the chiral and $CP_N$ models would then have additional
parameters. However, in our DJPT model, there is no nontrivial
vacuum parameters as  our scalar field $\phi$ vanishes at
infinity.  For a single fundamental soliton on $U(N)$ theory would
have not only position moduli parameters, but also scaling and
internal orientation paramters. When there are several solitons
presents, there would be relative size and orientation paramters
also. It would be interesting to find out the right counting.

\section{ Toda-Model on Noncommutative Plane}

On commutative plane, a class of  Chern-Simons soliton solution
has been associated with  solutions of the $SU(N)$
Toda-model\cite{Dunne:1990qe,Dunne:1992hq}. Especially for $N=2$,
the Toda-model is equivalent to the Liouville model.  On
noncommutative plane, the integrable generalization of $U(N)$
Toda-equation has not been stated  as far as we know, even though
there are several  works on the generalization of integrable
models on noncommutative plane.  We will show here that a
straightforward  extension of the work in
Ref.~\cite{Dunne:1990qe,Dunne:1992hq}, leads to  a generalization
of integrable Liouville and Toda-models on noncommutative and a
large class of its solutions. By examiming some simple example, we
will see this generalization is still highly nontrivial.

As on commutative plane, we start with an consistent ansatz for
the field configuration,
\be A_+= {\rm diag} (E_1,E_2,...,E_N) , \;\;\;\;  (\phi)_{ab} =
\delta_{a,b-1} h_a  \,\, (a=1,2,...N-1) \label{todaansatz}\ee
with $N$ components  $E_a$ and $N-1$ components $h_a$.  The
self-dual equations (\ref{gauss1}) and (\ref{selfdual}) on
noncommutative plane for this ansatz  become
\ba &&  \partial_+ h_a + E_ah_a -h_a E_{a+1}= 0 \;\;
(a=1,2,...,N-1)
\label{todaI1}  \\
&& \partial_+ (-E_1^\dagger) -\partial_- E_1 + [(-E_1^\dagger),
E_1] = h_1h_1^\dagger  \nonumber \\
 && \partial_+ (-E_a^\dagger) -\partial_- E_a +
[(-E_a^\dagger), E_a] = -h_{a-1}^\dagger h_{a-1} +h_{a}
h_{a}^\dagger\, \;(a=2,...,N-1) \nonumber \\
&&   \partial_+ (-E_N^\dagger) -\partial_- E_N + [(-E_N^\dagger),
E_N] = -h_{N-1}^\dagger h_{N-1} \label{todaI2} \ea
On noncommutative space the gauge field $A_+$ is not traceless and
so one could not solve the first set of equations (\ref{todaI1})
for $E_a$ to reduce the above coupled linear equations to the
second order equations for $h_a$ only. One has to live with these
coupled first order equations  (\ref{todaI1}) and (\ref{todaI2})
for $E_a$ and $h_a$ and regard them to define the $U(N)$ Toda
model on noncommutative plane. For $N=2$, the above equations
become the generalization of the Liouville equation on
noncommutative plane.  On commutative such reduction can be done
and one ends up with the second order equations for $\rho_a =
|h_a|^2$ which define the Toda model.

To find the solution of the above equations on noncommutative
plane, let us start with a set of explicit solutions, the
so-called `Toda-type', in terms of unitons on commutative
plan~\cite{Dunne:1992hq}. we will show that its  main gist   works
out  on noncommutative plane as well with due care on ordering of
operators. To find  for the solutions in $U(N)$ theory, we start
with $N$ dimensional vector $u$ such that
\be u^T = \left(  f_1(z) ,  f_2(z), ... f_N(z) \right)
\label{ueq}\ee
where $f_j(z)$ are `polynomials' of $z$. Now one defines $M_k$
which is a $N\times k$ matrix
\be M_k = (u, \partial_- u, \partial_-^2u, ..., \partial_-^{k-1}
u) \label{mkeq} \ee

There are many parameters characterize these polynomials, some of
them are moduli parameters of solitons and some of them are
redundant. As the moduli parameters change, the configurations can
be degenerated. As we have seen in the previous section, solitons
cannot collapse and disappear on noncommutative plane. They can
still spread all over space, leaving zero energy density. For
generic values of the parameters, there would be no common factors
among the polynomials $f_j(z)$.

For $k\le N$, one can define a hermitian projection operator
\be p_k = M_k (\bar{M}_k M_k)^{-1} \bar{M}_k \label{pkdefeq} \ee
which is a uniton with respect to an identity operator.  The
chiral energy for the group element $h=2p_k-1$ would be the
masimum degree of the polynomials $f_j(z)$.

We are now ready to adapt the results in
Ref.~\cite{Dunne:1990qe,Dunne:1992hq} on the relation between
Toda-model and the uniton method on commutative plane   to that on
noncommutative plane. Especially we generalize a theorem in
Ref.~\cite{Dunne:1992hq} to   show that a certain class of the
solutions of the chiral model on noncommutative plane can be gauge
transformed to the solution of the generalized Toda model
(\ref{todaI1}) and (\ref{todaI2}) on noncommutative plane. As the
following statement is proven, we state it as a theorem.

\vskip 1em

\noindent{\bf Theorem III:} For a $N$-dimensional vector $u$ in
Eq.~(\ref{ueq}), we have defined $N\times k$ vector $M_k$ in
Eq.~(\ref{mkeq}) and a projection operator $p_k$ in
Eq.~(\ref{pkdefeq}). Then the following operator
\be h = (2p_1-1)(2p_2-1)...(2p_{N-1}-1) \label{htodaeq} \ee
is a solution of the $U(N)$  chiral model field equation on
noncommutative plane. So that $\phi=A_+=
\frac{1}{2}h^{-1}\partial_+h $ describes the Chern-Simons
solitons. Furthermore, there exists a unitary transformation $g$
which leads to $g^\dagger \phi g$ and $g^\dagger A_+ g + g^\dagger
\partial_+ g$ are in the Toda ansatz (\ref{todaansatz}), and so become
the solution of  the generalized $U(N)$ Toda
equations (\ref{todaI1}) and (\ref{todaI2}) on noncommutative
plane.

\vskip 1em

\noindent {\bf Proof}: The linear algebra with noncommuting
variables are full of pit falls. In our case we will see that our
scope is just so  that some key lore of linear algebra works out
fine. The vector $u$ in Eq.~(\ref{ueq}) are given by $N$
polynomials of $z$ only. Thus as a function of $z$ we can discuss
the linear independence. We assume that the $N$ column vectors
$u,\partial_- u, ..., \partial_-^{N-1} u$ which all depend only on
$z$  are linearly independent. (If the degree of $d(u)$ is larger
than $N$ and polynomials are generic, then it is so.)  We consider
a  space ${\cal V}$ made of vectors given as   a linear
combination,
\be V = \sum_{r=1}^N (\partial_{-}^{r-1} u )\; c_r \ee
where $c_r(z,\bar{z})$'s are  scalar functions of $z$ and
$\bar{z}$. One start  with a unit column vector
\be e_1 = u \;(\bar{u}u)^{-1/2} \ee
in space ${\cal V}$, which shows that $p_1 = e_1 e_1^\dagger $.

As $p_2$ of Eq.~(\ref{pkdefeq}) is a hermitian $N\times N$
projection operators such that $(p_2)^2=p_2$, we can see that $p_2
u=u, p_2 \partial_- u= \partial_- u$. Now we  define another unit
column vector
\be e_2 = (1-p_1)\partial_- u \;\bigl(\partial_+\bar{u}
(1-p_1)\partial_-u\bigr)^{-1/2} \ee
in ${\cal V}$,  which is orthogonal to $e_1$ as $\bar{e}_1 e_2=0$.
Thus we see that $p_2 e_1=e_1, p_2 e_2=e_2$. Now we say
$q_2=e_1\bar{e}_1+e_2\bar{e}_2$, then we see $p_2q_2= q_2=p_2$ on
noncommutative space  as on commutative space.

By the Gramm-Schmidt process, we choose $e_r$ $(r\ge 2)$ to be a
unit vector,
\be e_r = (1-p_{r-1})\partial_-^{r-1} u\biggl(
\partial_+^{r-1} u (1-p_{r-1})\partial_-^{r-1}
u\biggr)^{-\frac{1}{2}} \ee
which is orthogonal to $e_1,...,e_{r-1}$. Again we see that $p_r=
e_1\bar{e}_1+...e_r\bar{e}_r$. Now  the unitary matrix defined as
\be g= (e_1,e_2,...,e_N) \ee
such that $g^{-1} = g^\dagger$  diagonalizes all the $p_r$'s so
that
\be g^{-1} p_r g = {\rm diag}(1,1,..,1,0,...,0) \ee
where $r+1$ entry is the first zero element.  Similar to the case
on the commutative plane, one can see that
\be \bar{e}_r \partial_+ e_s \neq  0  \;\; {\rm only}\;\;{\rm
for}\;\;  (s=r\;\; {\rm or} \;\; s=r+1) \ee
With $g^{-1}(\partial_+p_r)g= [g^{-1}\partial_+g,g^{-1}p_r g]$,
the above relation implies that
\be [g^{-1} (\partial_+ p_r) g]_{ab} =
-\delta_{a,r}\delta_{b,{r+1}}\bar{e}_r\partial_+ e_{r+1} \ee
Using the above observation, one can see that the group element
$h$ of Eq.~(\ref{htodaeq}) becomes
\be A_+= \phi = \frac{1}{2} h^{-1}\partial_+ h =
\sum_{r=1}^{N-1}\partial_+ p_r. \ee
which is describes the Chern-Simons solitons. Under the gauge
transformation by $g$, the scalar field $g^{-1}\phi g$ and  the
gauge field $g^{-1} A_+g + g^{-1}
\partial_+ g$ take the form of the Toda-ansatz (\ref{todaansatz}),
\ba &&  h_a = [g^{-1}\phi g ]_{ab}  = - \delta_{a,b-1} \;
e_a^\dagger
\partial_+ e_{a+1}, \;\;\; (a=1,2,...,N-1) \label{todaII1}\\
&&   E_a= g^{-1} A_+ g + g^{-1}
\partial_+ g = {\rm diag} (e_1^\dagger
\partial_+ e_1,e_2^\dagger \partial_+ e_2,...,e_N^\dagger \partial_+ e_N)
\label{todaII2} \ea
This completes the proof of the above theorem for the Toda-model
on noncommutative space.

Let us examine more closely the Toda-model and the Chern-Simons
solitons in the $U(2)$ case. The above proof provides  a class of
the general solution start with any two polynormials $f_1(z)$ and
$f_2(z)$ for the $u$ vector (\ref{ueq}). To feel this general
solution, let us examine the simplest ansatz,
\be u= \left(\begin{array}{c} z \\ c \end{array}\right) \ee
with a constant parameter $c$.  (For the rest of discussion, we
put $\theta=2$ and so $z=a$ for simplicity.) The corresponding
orthonormal vectors are
\ba && e_1 = \left(\begin{array}{c} z \\ c \end{array}\right)
\sqrt{\frac{1}{ \CN + |c|^2}}, \\
 && e_2= \left(\begin{array}{c} \bar{c} \\ -\bar{z}
\end{array}\right) \frac{1}{\bar{c}}\sqrt{\frac{|c|^2}{\CN+1+|c|^2} }  \ea
where $\CN=\bar{z}z$ is the number operator on the Hilbert space.
With the noncommtative parameter $\theta=2$ for the simplicity,
the scalar field (\ref{todaII1}) becomes
\be h_1= [g^{-1}\phi g]_{12} =
\frac{\sqrt{|c|^2}}{\sqrt{(\CN+|c|^2)(\CN+1+|c|^2)}}\ee
and the corresponding gauge field (\ref{todaII2}) becomes
 \ba &&  E_1=(g^{-1} A_+ g + g^{-1})_{11} = \left(
 \sqrt{\frac{\CN+|c|^2}{\CN+1+|c|^2}}- 1\right)a \\
 && E_2=(g^{-1} A_+ g + g^{-1})_{22} = \left(
 \sqrt{\frac{\CN+2+|c|^2}{\CN+1+|c|^2}}- 1 \right) a \ea
Note that the field strength is diagonal and becomes
\be [g^{-1}(F_{+-})g]_{11} = -[g^{-1}(F_{+-})g]_{22} =
\frac{|c|^2}{(\CN+|c|^2)(\CN+1+|c|^2)} \ee
This is the simplest solution of the $U(2)$ Toda-model or
Liouville model on noncommutative plane. Of course we can consider
more general solutions, which is quite straightforward.

As noted in Ref.~\cite{Dunne:1990qe}, one can have  a consistent
ansatz which is a slightly more general than the ansatz
(\ref{todaansatz}) for the Toda model. The gauge field in the new
ansatz is diagonal as  before. The scalar field which had nonzero
component only along simple roots of $SU(N)$ has  an additional
nonzero component, which  corresponds  to the lowest negative
root. This ansatz for the affine Toda model works as well on
noncommutative plane. The ansatz for $U(N)$ affine Toda model on
noncommutative plane is
\ba && A_+= {\rm diag} (E_1,E_2,...,E_N) , \nonumber \\
&&  (\phi)_{ab} = \delta_{a,b-1} h_a  \,\, (a=1,2,...N-1), \;\;
{\rm except \;\; for}\;\;  (\phi)_{N1} = h_N
\label{atodaansatz}\ea
which now have complex $N$ components  $E_a$ and complex $N$
components $h_a$. The self-dual equations (\ref{gauss1}) and
(\ref{selfdual}) on noncommutative plane for this ansatz  become
\ba &&  \partial_+ h_a + E_ah_a -h_a E_{a+1}= 0 \;\;
(a=1,2,...,N-1) \nonumber \\
&& \partial_+ h_N + E_N h_N- h_N E_1 = 0
\label{atodaI1}  \\
&& \partial_+ (-E_1^\dagger) -\partial_- E_1 + [(-E_1^\dagger),
E_1] = h_1h_1^\dagger  -h_N^\dagger h_N \nonumber \\
 && \partial_+ (-E_a^\dagger) -\partial_- E_a +
[(-E_a^\dagger), E_a] = -h_{a-1}^\dagger h_{a-1} +h_{a}
h_{a}^\dagger\, \;(a=2,...,N-1) \nonumber \\
&&   \partial_+ (-E_N^\dagger) -\partial_- E_N + [(-E_N^\dagger),
E_N] = -h_{N-1}^\dagger h_{N-1} +h_N h_N^\dagger \label{atodaI2}
\ea
For $N=2$, the above model on commutative space with a further
restriction on the solution reduces to sinh-Gordon equation. On
noncommutative plane, we cannot solve for $E_a$ in terms of $h_a$
and so we cannot get second order equations  for $h_a$ only. On
commutative plane, the nontrivial solution of the affine
Toda-equation has infinite charge $Q$~\cite{Dunne:1990qe}. It
would be very interesting to find some solutions of our
generalization of affine Toda-equation on noncommutative plane.

Our generalization of (affine) Toda, Liouville, sinh-Gordon model
on noncommutative plane is done directly on the gauge fields and
scalar field on noncommutative plane, resulting in coupled first
order equations.  Thus, our work contrasts to recent works on
integrable models on noncommutative plane, where the coupled first
order equations are reduced to the coupled second order equations,
which is then generalized to noncommutative space. While the later
has an appealing feature of having less number of variables, ours
has a feature that some of solutions can be obtained explicitly.
It remains to been seen whether the two procedures are compatible,
and whether the integrability survives in both cases.

\vskip 1em

\section{ Moduli Dynamics} \vskip 1em

The moduli space dynamics of chiral field would be
\be {\cal K}_{\rm chiral} = - \frac{1}{2}\int d^2x\;
\tr(h^{-1}\dot{h})^2 \ee
For the simple case $h=2p-1$, we get
\be {\cal K}_{p^2=p} = 2\int d^2x\; \tr(\dot{p})^2 \ee
which can be regarded as the soliton dynamics of the purely scalar
field on noncommutative plane in large noncommutative limit.
Finally, our Chern-Simons theory is first order Lagrangian. There
are some studies of these moduli
space~\cite{Gopakumar:2001yw,Lindstrom:2000kh,Lechtenfeld:2001uq,Furuta:2002ty}.
It provide the metric for the moduli space parameters.

When one consider the moduli dynamics of solitons in the chiral
model or the simpler version, $CP^N$ model,  some of the
parameters of solitons  has infinite inertia, or kinetic mess,
which implies that those parameters characterizes the vacuum where
solitons exist, not the moduli parameters of solitons. The
parameters of finite inertia can change with time.  The moduli
space of solitons on noncommutative space seems complete contrast
to that on commutative space. The reason is that the
noncommutativity of space provides a  short distance cut-off and
so solitons cannot collapse and disappear.

For Chern-Simons solitons  the selfdual configuration satisfy a
relation (\ref{aphieq}) in a gauge.  As they carry zero energy,
the Lagrangian (\ref{Lagrang}) becomes
\be {\cal K}_{\rm CS} = \frac{\nn i}{2\pi} \int d^2x \; \tr
\phi^\dagger \dot{\phi} \label{forderlag} \ee
modulo the Gauss law (\ref{gauss1}).  Clearly this is a first
order in time-derivative and so does not provide any metric on
moduli space.

We consider a single uniton solution (\ref{sunitonp}) in $U(N)$.
We notice that
\ba && \partial_+\partial_- p  =
-M\frac{1}{\bar{M}M}\partial_+\bar{M}(1-p)\partial_-M\frac{1}{\bar{M}M}
\bar{M} + (1-p)
\partial_-M\frac{1}{\bar{M}M}\partial_+\bar{M}(1-p) \nonumber \\
&& \partial_tp = (1-p) \partial_t M \frac{1}{\bar{M}M} \bar{M}\ea
The above first order Lagrangian (\ref{forderlag}) for the moduli
parameters for a single uniton  (\ref{sunitonp})  becomes
\be {\cal K}_{\rm CS} =  i\nn\; \theta \Tr \; \tr\;\biggl[
(\partial_+\partial_- p )\partial_t p \biggr] = 0 \ee
after integration by part.

The vanishing of the first order Lagrangian for the moduli space
of a single uniton which represents many fundamental Chern-Simons
solitons suggests  that the first order Lagrangian probably
vanishes for all possible self-dual configurations given by
Conjecture I.  While this  remains to be seen,  the first order
moduli space dynamics between Chern-Simons solitons are expected
to be trivial on general ground.

The total  angular momentum of many  fundamental Chern-Simons
solitons described by a single uniton is the sum of that for
individual ones and so independent of their moduli parameters. In
addition, the fundamental Chern-Simons solitons have integer spin
$\nn$ upon quantization. Thus they need no first order interaction
in short or long distance for the spin-statistics theorem to hold.
This is quite different from the behavior of two anyonic solitons,
each with spin $s$, whose classical angular momentum interpolates
between $2s$ in  large separation to $s^2$ in no separation. For
anyonic solitons one needs nontrivial first order kinetic terms
for  the statistical interaction.

\vskip 1.4em

 \centerline{\bf Acknowlegements}

\vskip 0.4em

 I appreciate the encouragements from Gerald Dunne,
Dongsu Bak and Choonkyu Lee. This work is supported in part by NSF
under Grant No. 0071512 and by grant No. R01-2003-000-10391-0 from
the Basic Research Program of the Korea Science \& Engineering
Foundation.

 \vfil


\newpage

\begin{thebibliography}{99} \frenchspacing


%%%%%%%%% review  of physics on noncommutative plane

%\cite{Douglas:2001ba}
\bibitem{Douglas:2001ba}
M.~R.~Douglas and N.~A.~Nekrasov,
%``Noncommutative field theory,''
Rev.\ Mod.\ Phys.\  {\bf 73} (2001) 977 [arXiv:hep-th/0106048].
%%CITATION = HEP-TH 0106048;%%



%\cite{Harvey:2001yn}
\bibitem{Harvey:2001yn}
J.~A.~Harvey,
%``Komaba lectures on noncommutative solitons and D-branes,''
arXiv:hep-th/0102076.
%%CITATION = HEP-TH 0102076;%%



%\cite{Szabo:2001kg}
\bibitem{Szabo:2001kg}
R.~J.~Szabo,
%``Quantum field theory on noncommutative spaces,''
Phys.\ Rept.\  {\bf 378} (2003) 207 [arXiv:hep-th/0109162].
%%CITATION = HEP-TH 0109162;%%

%%%%%%%% GMS solitons on noncommutative plane



%\cite{Gopakumar:2000zd}
\bibitem{Gopakumar:2000zd}
R.~Gopakumar, S.~Minwalla and A.~Strominger,
%``Noncommutative solitons,''
JHEP {\bf 0005} (2000) 020 [arXiv:hep-th/0003160].
%%CITATION = HEP-TH 0003160;%%

%%%%%%%%%%%%%%%%%%%%%%%%%%%%%%%

%\cite{Nekrasov:1998ss}
\bibitem{Nekrasov:1998ss}
N.~Nekrasov and A.~Schwarz,
 %``Instantons on noncommutative R**4 and (2,0) superconformal six  dimensional
%theory,''
Commun.\ Math.\ Phys.\  {\bf 198} (1998) 689
[arXiv:hep-th/9802068].
%%CITATION = HEP-TH 9802068;%%


%%%%%%%%%%%% Chern-Simons Solitons


%\cite{Dunne:1990qe}
\bibitem{Dunne:1990qe}
G.~V.~Dunne, R.~Jackiw, S.~Y.~Pi and C.~A.~Trugenberger,
%``Selfdual Chern-Simons Solitons And Two-Dimensional Nonlinear Equations,''
Phys.\ Rev.\ D {\bf 43}, 1332 (1991) [Erratum-ibid.\ D {\bf 45},
3012 (1992)].
%%CITATION = PHRVA,D43,1332;%%

%\cite{Dunne:1992hq}
\bibitem{Dunne:1992hq}
G.~V.~Dunne,
%``Chern-Simons solitons, toda theories and the chiral model,''
Commun.\ Math.\ Phys.\  {\bf 150}, 519 (1992)
[arXiv:hep-th/9204056].
%%CITATION = HEP-TH 9204056;%%


%%%%% unitons



\bibitem{uhlenbeck} K. Uhlenbeck, `` Harmonic Maps into Lie
Groups (Classical Solutions of the Chiral Model,'' Jour. Diff.
Geom. {\bf 30} (1989) 1.

\bibitem{wood} J.C. Wood, ``Explicit Construction and
Parameterization of Harmonic Two Sphers in the Unitary Group,''
Proc. London Math. Soc. {\bf 58} (1989) 608.




%%%%%%%%%Chern-Simons review

%\cite{Dunne:ai}
\bibitem{Dunne:ai}
G.~V.~Dunne,
%``Selfdual Chern-Simons Theories,''
Lect.\ Notes Phys.\  {\bf M36} (1995) 1;
%%CITATION = LNPHA,M36,1;%%
K.~M.~Lee,
%``Relativistic self-dual Chern-Simons systems: A perspective,''
Int.\ J.\ Mod.\ Phys.\ A {\bf 12} (1997) 1003
[arXiv:hep-th/9609137].
%%CITATION = HEP-TH 9609137;%%

%%%%%% ``Ward Conjecture''
\bibitem{ward2} R.S. Ward, Phil. Trans. Roy. Soc. Lond. {\bf A}
315 (1985) 451; {\it Multidimensional integrable systems}, Lect.
Notes. Phys. {\bf 280} (Springer, 1986) 106, {\it Integrable
systems in twistor theory}, in ``Twistes in Mathematics and
Physics'' (Cambridge UP, 1990) 246.

%%%%%%%%%%%%%%%%%%%%%%%%%%%%%%% GMS solitons Dynamics

%\cite{Gopakumar:2001yw}
\bibitem{Gopakumar:2001yw}
R.~Gopakumar, M.~Headrick and M.~Spradlin,
%``On noncommutative multi-solitons,''
Commun.\ Math.\ Phys.\  {\bf 233} (2003) 355
[arXiv:hep-th/0103256].
%%CITATION = HEP-TH 0103256;%%


%\cite{Lindstrom:2000kh}
\bibitem{Lindstrom:2000kh}
U.~Lindstrom, M.~Rocek and R.~von Unge,
%``Non-commutative soliton scattering,''
JHEP {\bf 0012} (2000) 004  [arXiv:hep-th/0008108];
%%CITATION = HEP-TH 0008108;%%
L.~Hadasz, U.~Lindstrom, M.~Rocek and R.~von Unge,
%``Noncommutative multisolitons: Moduli spaces, quantization, finite Theta
%effects and stability,''
JHEP {\bf 0106} (2001) 040  [arXiv:hep-th/0104017].
%%CITATION = HEP-TH 0104017;%%




%\cite{Lechtenfeld:2001uq}
\bibitem{Lechtenfeld:2001uq}
O.~Lechtenfeld, A.~D.~Popov and B.~Spendig,
%``Noncommutative solitons in open N = 2 string theory,''
JHEP {\bf 0106} (2001) 011 [arXiv:hep-th/0103196];
%%CITATION = HEP-TH 0103196;%%
O.~Lechtenfeld and A.~D.~Popov,
%``Noncommutative multi-solitons in 2+1 dimensions,''
JHEP {\bf 0111}, 040 (2001) [arXiv:hep-th/0106213];
%%CITATION = HEP-TH 0106213;%%
O.~Lechtenfeld and A.~D.~Popov,
%``Scattering of noncommutative solitons in 2+1 dimensions,''
Phys.\ Lett.\ B {\bf 523}, 178 (2001) [arXiv:hep-th/0108118].
%%CITATION = HEP-TH 0108118;%%


%%%%%%%%%%% CP(N) model on noncommutative plane


%\cite{Lee:2000ey}
\bibitem{Lee:2000ey}
B.~H.~Lee, K.~M.~Lee and H.~S.~Yang,
%``The CP(n) model on noncommutative plane,''
Phys.\ Lett.\ B {\bf 498}, 277 (2001) [arXiv:hep-th/0007140].
%%CITATION = HEP-TH 0007140;%%

%\cite{Furuta:2002ty}
\bibitem{Furuta:2002ty}
K.~Furuta, T.~Inami, H.~Nakajima and M.~Yamamoto,
%``Low-energy dynamics of noncommutative CP(1) solitons in 2+1 dimensions,''
Phys.\ Lett.\ B {\bf 537} (2002) 165 [arXiv:hep-th/0203125].
%%CITATION = HEP-TH 0203125;%%



%\cite{Otsu:2003fq}
\bibitem{Otsu:2003fq}
H.~Otsu, T.~Sato, H.~Ikemori and S.~Kitakado,
%``New BPS solitons in 2+1 dimensional noncommutative CP(1) model,''
JHEP {\bf 0307} (2003) 054 [arXiv:hep-th/0303090];
%%CITATION = HEP-TH 0303090;%%
H.~Otsu, T.~Sato, H.~Ikemori and S.~Kitakado,
%``Lost equivalence of nonlinear sigma and CP(1) models on noncommutative
%space,''
arXiv:hep-th/0404140.
%%CITATION = HEP-TH 0404140;%%

%%%%%%%%%% Chern-Simons level quantization

%\cite{Deser:vy}
\bibitem{Deser:vy}
S.~Deser, R.~Jackiw and S.~Templeton,
%``Three-Dimensional Massive Gauge Theories,''
Phys.\ Rev.\ Lett.\  {\bf 48} (1982) 975;
%%CITATION = PRLTA,48,975;%%


%\cite{Bak:2001ze}
\bibitem{Bak:2001ze}
D.~Bak, K.~M.~Lee and J.~H.~Park,
%``Chern-Simons theories on noncommutative plane,''
Phys.\ Rev.\ Lett.\  {\bf 87} (2001) 030402
[arXiv:hep-th/0102188].
%%CITATION = HEP-TH 0102188;%%


%\cite{Nair:2001rt}
\bibitem{Nair:2001rt}
V.~P.~Nair and A.~P.~Polychronakos,
%``On level quantization for the noncommutative Chern-Simons theory,''
Phys.\ Rev.\ Lett.\  {\bf 87} (2001) 030403
[arXiv:hep-th/0102181].
%%CITATION = HEP-TH 0102181;%%


%%%%%%%%%U(N) and U(1) universality

%\cite{Gross:2000ss}
%\bibitem{Gross:2000ss}
%D.~J.~Gross and N.~A.~Nekrasov,
%``Solitons in noncommutative gauge theory,''
%JHEP {\bf 0103} (2001) 044 [arXiv:hep-th/0010090].
%%CITATION = HEP-TH 0010090;%%

%\cite{Bak:2000zm}
%\bibitem{Bak:2000zm}
%D.~Bak, K.~M.~Lee and J.~H.~Park,
%``Comments on noncommutative gauge theories,''
%Phys.\ Lett.\ B {\bf 501} (2001) 305 [arXiv:hep-th/0011244].
%%CITATION = HEP-TH 0011244;%%


%%%%%%%% Integrable Models on noncommutative space

%\cite{integrable}
\bibitem{integrable}
A.~Dimakis and F.~Mueller-Hoissen,
%``Bicomplexes, integrable models, and noncommutative geometry,''
Int.\ J.\ Mod.\ Phys.\ B {\bf 14} (2000) 2455
[arXiv:hep-th/0006005];
%%CITATION = HEP-TH 0006005;%%
I.~Cabrera-Carnero and M.~Moriconi,
%``Noncommutative integrable field theories in 2d,''
Nucl.\ Phys.\ B {\bf 673} (2003) 437 [arXiv:hep-th/0211193];
%%CITATION = HEP-TH 0211193;%%
M.~Moriconi and I.~Cabrera-Carnero,
%``Noncommutative field theories and integrable models in 2d,''
arXiv:hep-th/0303168.
%%CITATION = HEP-TH 0303168;%%
M.~Hamanaka and K.~Toda,
%``Towards noncommutative integrable systems,''
Phys.\ Lett.\ A {\bf 316} (2003) 77 [arXiv:hep-th/0211148];
%%CITATION = HEP-TH 0211148;%%
M.~Hamanaka and K.~Toda,
%``Noncommutative Burgers equation,''
J.\ Phys.\ A {\bf 36} (2003) 11981 [arXiv:hep-th/0301213];
%%CITATION = HEP-TH 0301213;%%
M.~Hamanaka,
%``Commuting flows and conservation laws for noncommutative Lax hierarchies,''
arXiv:hep-th/0311206;
%%CITATION = HEP-TH 0311206;%%
M.~T.~Grisaru and S.~Penati,
%``The noncommutative sine-Gordon system,''
arXiv:hep-th/0112246;
%%CITATION = HEP-TH 0112246;%%
M.~T.~Grisaru, L.~Mazzanti, S.~Penati and L.~Tamassia,
%``Some properties of the integrable noncommutative sine-Gordon system,''
JHEP {\bf 0404} (2004) 057 [arXiv:hep-th/0310214];
%%CITATION = HEP-TH 0310214;%%
M. Legar\'e,
%{\it Reduced systems of (2,2) pseudo-Euclidean
%noncommutative self-dual Yang-Mills theories},
J. Phys. A: Math.Gen. {\bf 35} (2002) 5489.



%%%%%%%%%%%%%%%%%%

%%%%%%%%%%%%%%% More recent works on CP_N and Chiral Model



%\cite{Wolf:2002jw}
\bibitem{Wolf:2002jw}
M.~Wolf,
 %``Soliton antisoliton scattering configurations in a noncommutative sigma
%model in 2+1 dimensions,''
JHEP {\bf 0206} (2002) 055 [arXiv:hep-th/0204185];
%%CITATION = HEP-TH 0204185;%%
J.~Murugan and R.~Adams,
%``Comments on noncommutative sigma models,''
JHEP {\bf 0212} (2002) 073 [arXiv:hep-th/0211171].
%%CITATION = HEP-TH 0211171;%%

%%%%%%%%%%%%%%%%%%%%%%


%


%%%%%%%%%%%%%%%%% Chern-Simons vortex on noncommutative plane

%\cite{Bak:2001sg}
\bibitem{Bak:2001sg}
D.~Bak, S.~K.~Kim, K.~S.~Soh and J.~H.~Yee,
%``Noncommutative Chern-Simons solitons,''
Phys.\ Rev.\ D {\bf 64} (2001) 025018 [arXiv:hep-th/0102137].
%%CITATION = HEP-TH 0102137;%%

%\cite{Hadasz:2003yx}
\bibitem{Hadasz:2003yx}
L.~Hadasz, U.~Lindstrom, M.~Rocek and R.~von Unge,
%``Time dependent solitons of noncommutative Chern-Simons theory coupled to
%scalar fields,''
arXiv:hep-th/0309015;
%%CITATION = HEP-TH 0309015;%%
P.~A.~Horvathy and P.~C.~Stichel,
%``Moving vortices in noncommutative gauge theory,''
Phys.\ Lett.\ B {\bf 583} (2004) 353 [arXiv:hep-th/0311157].
%%CITATION = HEP-TH 0311157;%%


%%%%%%%%%%%%%%%%%%%%%%%%%%%%%%%%%%%%%%%%%%%%%%%%%%%%


%%%%%%%%%%%%% chiral models and unitons on commutative plane



\bibitem{valli} G. Valli, {\it On the energy spectrum of harmonic
two shperes in unitary groups}, Topology {\bf 27} (1988) 129.


\bibitem{zak} B. Piette and W. Zakrzewski, `` General Solutions of
the U(3) and U(4) Chiral Sigma Models in Two Dimensions,'' Nucl.
Phys. {\bf B300} (1988) 207; ``Some Classes of General Solutions
of the U(N) Chiral $\sigma$ Models in Two Dimensions, '' Journ.
Math. Phys. {\bf 30} (1989) 2233.

\bibitem{ward} R.S. Ward, {\it Classical solutions of the chiral
model, unitons and holomorphic vector bundles}, Comm. Math. Phys.
{\bf 128} (199) 319.





%%%%%%%%%%% relativistic Chern-Simons theories

%\cite{Jackiw:1990pr}
\bibitem{Jackiw:1990pr}
R.~Jackiw, K.~M.~Lee and E.~J.~Weinberg,
%``Selfdual Chern-Simons Solitons,''
Phys.\ Rev.\ D {\bf 42} (1990) 3488.
%%CITATION = PHRVA,D42,3488;%%


%\cite{Lee:1990bd}
\bibitem{Lee:1990bd}
K.~M.~Lee,
%``Relativistic Nonabelian Selfdual Chern-Simons Systems,''
Phys.\ Lett.\ B {\bf 255} (1991) 381.
%%CITATION = PHLTA,B255,381;%%

%\cite{Dunne:fs}
\bibitem{Dunne:fs}
G.~V.~Dunne,
%``Relativistic Selfdual Chern-Simons Vortices With Adjoint Coupling,''
Phys.\ Lett.\ B {\bf 324} (1994) 359.
%%CITATION = PHLTA,B324,359;%%



\end{thebibliography}
\end{document}